\documentclass[prl,twocolumn,a4paper]{revtex4}
\usepackage{graphicx,amsmath,bm}

\begin{document}

\title{Faithful entanglement swapping based on sum-frequency generation}
\date{\today}
\author{Nicolas Sangouard}
\affiliation{Group of Applied Physics, University of Geneva, 1211 Geneva 4, Switzerland}
\author{Bruno Sanguinetti}
\affiliation{Group of Applied Physics, University of Geneva, 1211 Geneva 4, Switzerland}
\author{No\'{e} Curtz}
\affiliation{Group of Applied Physics, University of Geneva, 1211 Geneva 4, Switzerland}
\author{Nicolas Gisin}
\affiliation{Group of Applied Physics, University of Geneva, 1211 Geneva 4, Switzerland}
\author{Rob Thew}
\affiliation{Group of Applied Physics, University of Geneva, 1211 Geneva 4, Switzerland}
\author{Hugo Zbinden}
\affiliation{Group of Applied Physics, University of Geneva, 1211 Geneva 4, Switzerland}

\begin{abstract}
We show that an entanglement swapping operation performed with spontaneous parametric down-conversion can be made faithful without post-selection using sum-frequency generation. This invites us to revisit the sum-frequency process and from a proof-of-principle experiment, we demonstrate that it provides a realistic solution for non-linear optics at the single-photon level. This opens the way to attractive alternatives to six-photon protocols based on linear optics used e.g. for the heralded creation of maximally entangled pairs or for device-independent quantum key distribution. 
\end{abstract}
\maketitle

\begin{figure}
\includegraphics[width=5.5 cm]{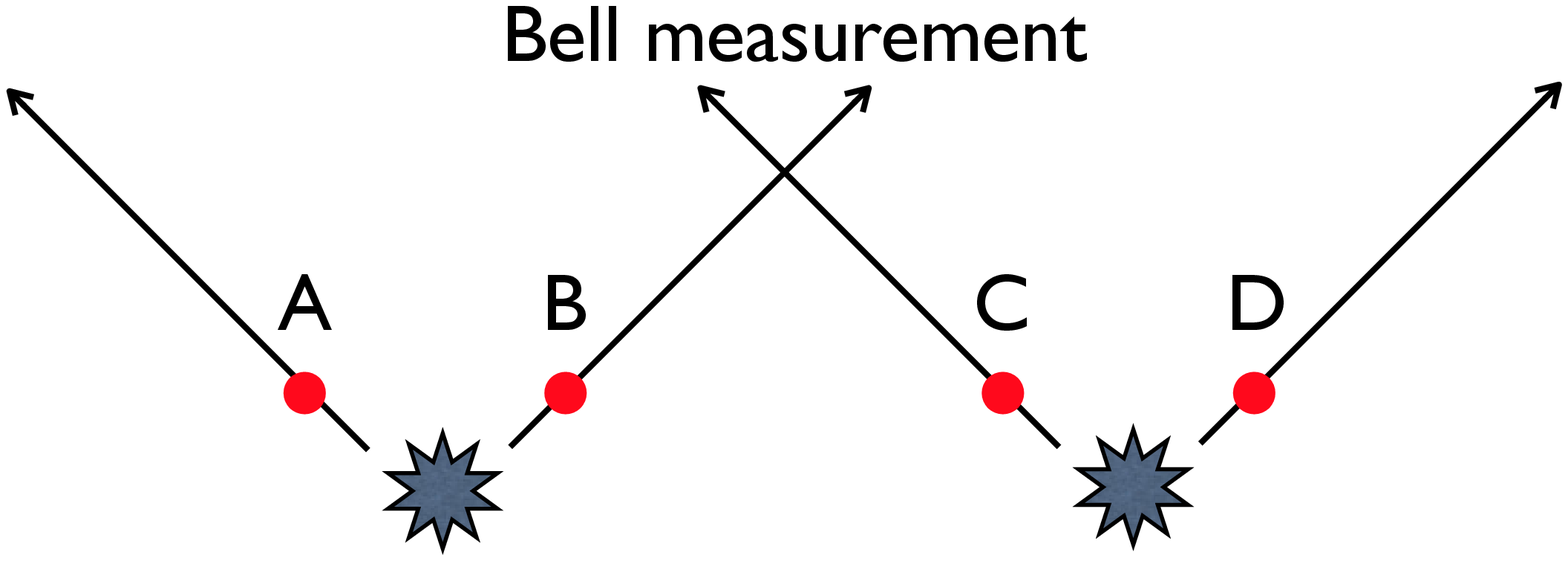}
\caption{Basic scheme for entanglement swapping. Sources (stars) produce entangled pairs, A-B and C-D. The projection of B and C in the Bell basis entangles the photons A and D.}
\label{fig1}
\end{figure}

\paragraph{Introduction}
A fascinating feature of entanglement is that it can be swapped \cite{Zukowski93}. Given two entangled photons, say the photons A and B, and another entangled pair C and D (see Fig.\,\ref{fig1}), it is possible to entangle A and D by performing a joint measurement of photons B and C in the Bell basis, provided that the result of the measurement is communicated to A and D. Hence, the two latter photons end up entangled even though they have never interacted.\\
The most natural approach to implement an entanglement swapping operation would be to use pair sources based on spontaneous parametric down-conversion (SPDC) and linear optical elements to perform a partial Bell measurement. This would allow one to realize a heralded entangled pair source - the success of the Bell measurement serving as the heralding event for the creation of an entangled pair. However, the emission of multi-pairs, inherent to SPDC, inevitably reduces the fidelity of the state conditioned on a successful Bell measurement, c.f. below. This makes the entanglement swapping protocol useless in practice without post-selection. Actually, it has been shown that the heralded production of a maximally entangled photon pair using only conventional SPDC sources, linear optics and projective measurements, requires at least three entangled pairs \cite{Kok00}, making the practical realization very challenging \cite{note_Barz10_Wagenknecht10}. Here, we show that the entanglement swapping operation with pairs coming from SPDC sources, can be made faithful if the Bell measurement uses sum-frequency generation (SFG). Such a process can effectively serve as a non-linear filter, inhibiting multi-pair emissions that would otherwise corrupt the Bell state measurement. SFG has already exhibited its usefulness in quantum information, e.g. to perform a complete Bell state discrimination \cite{Kim01}, to characterize broadband entangled photons \cite{Dayan05}, to highlight the shaping of entangled-photon waveform \cite{shapingsinglephoton} and for frequency conversion of single photons \cite{conver}. We demonstrate from a proof-of-principle experiment that the SFG in a non-linear waveguide with two broadband photons can be efficient enough to make our proposal attractive with respect to six-photon schemes that are required e.g. to herald the creation of maximally entangled pairs from SPDC sources \cite{Sliwa03}. Furthermore, since SFG can be driven by telecom wavelengths, our proposal can be used to herald the creation of entangled pairs at a distance. This offers a competitive alternative to linear-optics based solutions to overcome the problem of losses in device-independent quantum key distribution \cite{Gisin10}.\\

\paragraph{Failure of entanglement swapping with SPDC and linear optics} We first recall the principle of entanglement swapping with SPDC sources and linear optics, by choosing the time degree of freedom (the discussion would be similar for other degrees of freedom). In SPDC, a strong light pulse is sent through a non-linear crystal and with a small probability $p_{ab},$ a photon from the beam decays into two photons, one into the spatial mode $A,$ the other one into $B,$ such that energy and momentum are conserved. If the non-linear crystal is pumped by two coherent pulses \cite{Brendel99}, the photon pair is created in a coherent superposition of two modes, the early $e$ and late $\ell$ modes, with a delay corresponding to the one between the two pump pulses, i.e. the so-called time-bin entanglement. Taking the double-pair emission into account, the resulting state is well described by 
\begin{eqnarray}
\label{rhoab}
&&\rho_{ab} \approx |00\rangle\langle00| +p_{ab}\Big|\frac{a_e^\dagger b_e^\dagger-a_{\ell}^\dagger b_{\ell}^\dagger}{\sqrt{2}}\big\rangle\big\langle \frac{a_e^\dagger b_e^\dagger-a_{\ell}^\dagger b_{\ell}^\dagger}{\sqrt{2}}\Big| \\ \nonumber
&&+ \frac{3p_{ab}^2}{4} \Big|\frac{\left(a_e^\dagger b_e^\dagger-a_{\ell}^\dagger b_{\ell}^\dagger\right)^2}{2\sqrt{3}}\big\rangle\big\langle \frac{\left(a_e^\dagger b_e^\dagger-a_{\ell}^\dagger b_{\ell}^\dagger\right)^2}{2\sqrt{3}}\Big| + o(p_{ab}^3).
\end{eqnarray}
A second source, located far away from the previous one, is pumped simultaneously to produce another time-bin entangled pair in modes C and D with the probability $p_{cd}.$ The corresponding state $\rho_{cd}$ is similar to the state $\rho_{ab}$ with $a$ and $b$ replaced by $c$ and $d$ respectively. The modes B and C are then combined on a 50-50 beamsplitter. The modes after this beam splitter are $\delta_{e/\ell}=\frac{1}{\sqrt{2}}(b_{e/\ell}+c_{e/\ell})$ and $\bar \delta_{e/\ell}=\frac{1}{\sqrt{2}}(b_{e/\ell}-c_{e/\ell}).$ We are interested in the detection of two photons - for example one in $\delta_e$ and one in $\bar \delta_\ell.$ Let us consider separately the different contributions. If one pair is emitted from each of the sources, the expected detections can be produced with probability $\frac{1}{8}p_{ab}p_{cd}$ \cite{note} and the resulting state $\frac{1}{\sqrt{2}}(a_e^\dagger d_{\ell}^\dagger-a_{\ell}^\dagger d_e^\dagger)|0\rangle$ corresponds to the desired entanglement of remote photons A and D. However, the terms associated to the emission of two pairs from the same source may also produce a photon in $\delta_e$ and another one in $\bar \delta_\ell$ with probability  $\frac{1}{16}p_{ab}^2$ $(\frac{1}{16}p_{cd}^2).$ Such detections project the remaining modes A and D onto $a_e^\dagger a_\ell^\dagger |0\rangle$ ($d_e^\dagger d_\ell^\dagger |0\rangle$ respectively). Hence, neglecting the terms associated to the emission of more than two pairs, the conditional state shared by $A$ and $D$ 
\begin{eqnarray}
&\rho_{\text{swap}}=&\frac{1}{16}p_{ab}^2 \big|a_e^\dagger a_\ell^\dagger \big\rangle\big\langle a_e^\dagger a_\ell^\dagger \big|+\frac{1}{16} p_{cd}^2 \big|d_e^\dagger d_\ell^\dagger \big\rangle\big\langle d_e^\dagger d_\ell^\dagger \big|\\ 
\nonumber
&+ \frac{1}{8}p_{ab}p_{cd} & \big| \frac{1}{\sqrt{2}} \left(a_e^\dagger d_{\ell}^\dagger-a_{\ell}^\dagger d_e^\dagger \right) \big\rangle \big\langle \frac{1}{\sqrt{2}} \left(a_e^\dagger d_{\ell}^\dagger-a_{\ell}^\dagger d_e^\dagger \right)  \big|
\end{eqnarray}
is poorly entangled, i.e. its fidelity with respect to a maximally entangled state 
$
F=\big \langle  \frac{1}{\sqrt{2}} \left(a_e^\dagger d_{\ell}^\dagger-a_{\ell}^\dagger d_e^\dagger \right) \big | \rho_{\text{swap}} \big | \frac{1}{\sqrt{2}} \left(a_e^\dagger d_{\ell}^\dagger-a_{\ell}^\dagger d_e^\dagger \right) \big \rangle 
$
is upper bounded by $1/2.$ To mitigate the multi-pair problem, one usually post-selects the events corresponding to one detection at both ends (A and D). However, post-selection of this kind strongly limits the applications arising from the entanglement swapping operations. For example, they do not allow for the heralded creation of entangled pairs and they are incompatible with device-independent quantum key distribution since they inevitably open the detection loophole, c.f. below. \\

\begin{figure}
\includegraphics[width=5.5 cm]{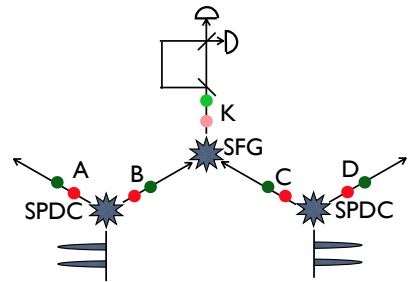}
\caption{Proposed setup for faithful entanglement swapping. The Bell measurement is based on SFG followed by an interferometer (made with one optical switch and one beamsplitter) to erase the which-path information, c.f. text for details.}
\label{fig2}
\end{figure}

\paragraph{Principle of our proposal} To overcome the limitation due to the multi-pair emission, we propose to use a non-linear process for the Bell measurement. Fig.\,\ref{fig2} shows the setup that we have in mind. The modes B and C, with well distinct frequencies, are combined into a non-linear medium at the central station. With a small probability, one photon in B combines with one photon in C to create a photon in K with a frequency corresponding to the sum of frequencies B and C. Note that if two input photons come from the same mode, they cannot create a photon in K due to energy conservation. This is at the heart of an attractive Bell measurement that we detail in what follows. Consider two temporal modes, the early and late modes as before. The SFG can generate a photon in K only if the input photons arrive at the same time. Hence, it is described by the Hamiltonian 
$
H=i \alpha (\kappa_e^\dagger b_e c_e - \kappa_\ell^\dagger b_\ell c_\ell) + h.c.
$ 
where $\kappa_e$ and $\kappa_\ell$ are associated to the photons created by the early and late modes respectively. The coupling constant $\alpha$ contains the material nonlinear susceptibility and the beam geometry. Let us start with $\rho_{ab}\otimes\rho_{cd}$ where $\rho_{ab}$ and $\rho_{cd}$ are as before (\ref{rhoab}). The dynamics is given by $\texttt{1}_{ad} \times e^{-iHt}$ where $\texttt{1}_{ad}$ is the identity for the modes A and D. Neglecting the terms associated with the emission of more than two pairs, one finds that the state resulting from a successful SFG is given by
$
\frac{1}{\sqrt{2}}\left(a_e^\dagger \kappa_e^\dagger d_e^\dagger-a_{\ell}^\dagger \kappa_{\ell}^\dagger
d_{\ell}^\dagger \right)|0\rangle.
$
To erase the information on the creation time, the modes $\kappa_e$ and $\kappa_\ell$ are then sent into an unbalanced Mach-Zehnder interferometer. The detection of one photon in $\frac{1}{\sqrt{2}}(|\kappa_e^\dagger\rangle+|\kappa_{\ell}^\dagger\rangle)$ projects the state of the remaining modes into 
$
\frac{1}{\sqrt{2}}\left(a_e^\dagger d_e^\dagger-a_{\ell}^\dagger d_{\ell}^\dagger \right)|0\rangle,
$
i.e. in a maximally entangled state. This offers interesting opportunities, e.g. to distribute entanglement at a distance in an heralded way, c.f. below. Taking the detection of one photon in $\frac{1}{\sqrt{2}}(|\kappa_e^\dagger\rangle-|\kappa_{\ell}^\dagger\rangle)$ into account, the success probability is given by $\frac{1}{2} \eta_{\text{SFG}} p_{ab} p_{cd}$ where $\eta_{\text{SFG}}=(\alpha \tau)^2$ is the SFG efficiency, $\tau$ being the characteristic interaction time between the photons and the non-linear medium. This provokes the natural question: is SFG efficient enough to be useful? \\

\paragraph{Heralded entangled-pair source} The setup presented in Fig.\,\ref{fig2} allows one to conditionally prepare a maximally entangled state of two photons -- the successful preparation being unambiguously heralded by the detection of a single photon resulting from the SFG. Alternative proposals associated with this endeavor but based on linear optics only, require the coincident detection of at least four auxiliary photons \cite{Kok00}. We focus on the proposal presented in Ref. \cite{Sliwa03} which is the only one performed experimentally \cite{note_Barz10_Wagenknecht10} so far. It uses the six photon component of the SPDC output. Specifically, a non-linear crystal produces two beams containing photons with pairwise correlated polarizations. A fourfold coincident detection performed on a fraction of the output beams, picked off with variable beamsplitters (which transmit a photon with a probability $\cos^2\theta$) leaves the remaining modes in a maximally entangled state. Taking the four-pair emission into account as well as the non-unit detection efficiency $\eta$ of the auxiliary photons, the fidelity of the heralded pair is well approximated by
$
F \approx \frac{\cos^4{\theta}}{(1-\eta\sin^2\theta)^2}\left(1-\frac{13}{4}p(1-\eta \sin^2\theta)^2\right)
$
in the limit where the success probability $p$ for the creation of one pair is small. The success probability for the fourfold coincidence is given by $P \approx \frac{1}{4}p^3\eta^4\sin^8\theta(1-\eta \sin^2\theta)^2(1+\frac{13}{4}p(1-\eta\sin^2\theta)^2)$ \cite{footnote}. Note that $\eta=\eta_c\eta_d$ contains the efficiency with which a photon is coupled into a mono-mode fiber $\eta_c$ and the detection efficiency $\eta_d.$ If we assume an overall detection efficiency $\eta=0.6,$ and if we require the fidelity of the entangled pairs to be $F  \geq 0.9,$ one finds the optimal values $p = 1.5 \times 10^{-2}$ and $\cos^2\theta = 0.93$ leading to $P \approx 3 \times 10^{-12}.$ For comparison with our proposal, we account for the non-unit coupling efficiency $\eta_c$ of modes B and C  and for the non-unit detection efficiency $\eta=\eta_c\eta_d$ of mode K. One finds that one entangled photon pair with fidelity $F\approx{1-3p}$ is heralded with probability $P\approx \frac{1}{2}\eta_c^2\eta\eta_\text{SFG} p^2(1+3p)$ where $p=p_{ab}=p_{cd}.$ For $\eta_c=\eta_d=\sqrt{0.6},$ $P \approx 3 \times 10^{-12}$ and $F \geq 0.9$ are achieved for $p \approx 3.3\times 10^{-2}$ provided that the efficiency of the SFG is $\eta_{\text{SFG}} \geq 1.4 \times 10^{-8}.$ Note that for lower detection efficiencies, for example $\eta=0.4,$ which is still higher than the overall detection efficiency reported in \cite{note_Barz10_Wagenknecht10}, $\eta_{\text{SFG}} \geq 3.1 \times 10^{-9}$ is sufficient for our proposal to achieve the same rate of entangled states (with $F\geq0.9$) than the one based solely on linear optics. \\

\paragraph{Potential application in QKD} An attractive feature of the swapping operation described above is that it can be performed at a distance. This is particularly useful to overcome the problem of transmission losses in device-independent quantum key distribution (DI-QKD) \cite{diqkd} where the secrecy  of the key relies solely on the violation of a Bell inequality. However, a necessary condition to insure the security in DI-QKD is to close the detection loophole. This is a major challenge in optical Bell tests,  since the detection efficiency, i.e. the product of the transmission efficiency (including the coupling into the fiber) and the photon-detector efficiency, required to rule out attacks based on the detection loophole is very high, typically larger than $82.8$\% for the CHSH inequality in the absence of other limitations. Even assuming perfect photo-detection and lossless components, the transmission efficiency of a 5~km long optical-fiber at telecom wavelength is roughly of 80\%. Transmission losses thus represent a fundamental limitation for the realization of a detection-loophole free Bell test on any distance relevant for QKD.\\
\begin{figure}
\includegraphics[width=7 cm]{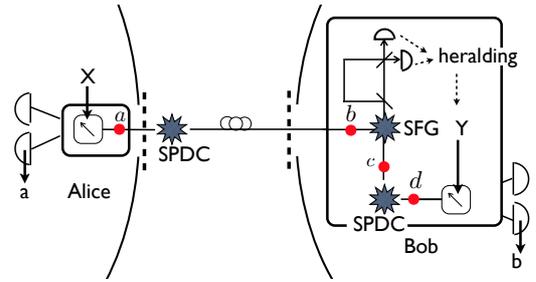}
\caption{Proposed setup to overcome the problem of transmission losses in DI-QKD. Each Alice and Bob boxes includes a measurement apparatus allowing them to choose a measurement (X and Y respectively) and to obtain outputs (a and b). If the resulting probability distribution $P(ab|XY)$ violates the CHSH inequality by a sufficient amount, a key can distill from the remaining data. Since Bob inputs a Y only when a SFG photon is detected, Alice and Bob can safely discard all events where a photon is lost during the transmission.}
\label{fig3}
\end{figure} 

The unique proposal allowing one to circumvent the problem of transmission losses requires three photon-pairs, linear optical elements and projective measurements \cite{Gisin10}. The performance of this scheme has been evaluated for a DI-QKD protocol based on the CHSH inequality \cite{QKD-DI-PRL}. Here, we focus on the case where the detectors are trusted. By considering a fiber attenuation of $0.2$\,dB/km, corresponding to telecom wavelength photons, a coupling efficiency of $\eta_c=0.9,$ a detection efficiency $\eta_d=0.8$ and sources with a repetition rate of $10$ Ghz, the proposal of Ref. \cite{Gisin10} achieves a rate of about $7$bits/min on a distance of 10~km. \\
Alternatively, SFG provides a simpler solution requiring only two pairs. The principle is shown in Fig.\,\ref{fig3}. An entangled pair source is located close to Alice's location. Each of Alice's and Bob's locations includes a measurement apparatus. Furthermore, Bob's box contains an entangled pair source and a non-linear crystal suitable for SFG. When a photon is detected after the SFG crystal, Bob knows that he shares an entangled state with Alice. Performing a measurement only when he gets a heralding signal, Alice and Bob can safely discard all the events where a photon got lost in the quantum channel. Consequently, the overall detection efficiency required to close the detection loophole does not depend anymore on the transmission efficiency but merely reduces to the intrinsic detection efficiency of Alice's and Bob's measurement apparatus. Taking the finite coupling and detection efficiency into account ($\eta_c=0.9$ and $\eta_d=0.8$ as before), one finds for a distance of 10~km, the optimal values $p_{ab} \approx p_{cd} = 3.7\times10^{-2}$ so that a key rate of $7$bits/min is achieved provided that the efficiency of SFG is $\eta_{\text{SFG}} \geq 6 \times 10^{-7}.$\\

\begin{figure}[htbp]
\includegraphics[width=6.5cm]{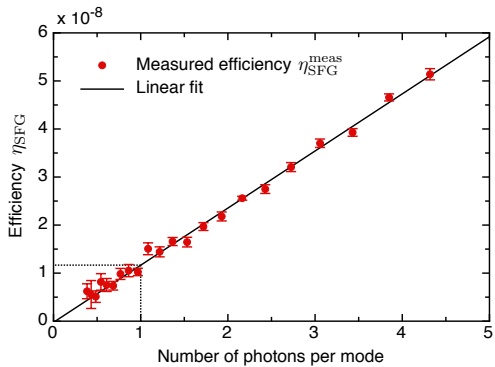}
\caption{Probability $\eta_\text{SFG}$ that a signal photon ($\lambda=1563\,\text{nm},\Delta\lambda=0.3\,\text{nm}$) is upconverted when interacting with a weak pump ($\lambda=1557\,\text{nm},\Delta\lambda=1.2\,\text{nm}$) inside the waveguide, plotted against the number of photons per mode (equal for pump and signal). Experimentally pump and signal photons are obtained by attenuating and filtering a Light Emitting Diode. Upconverted photons are separated from the residual pump using a prism, they are then detected with a single photon detector (IDQ ID100, 6\% efficiency). Dark counts (2.2\,Hz) have been subtracted, and injection (3dB), reflection (0.6 dB) and propagation (0.2 dB) losses have been taken into account.}
\label{fig4}
\end{figure}

\paragraph{Efficiency of SFG} Parametric processes at the single photon level have usually been considered too inefficient to be of practical interest. Nevertheless, recent experiments have put these processes to good use demonstrating the direct generation of photon triplets~\cite{Hubel10} and efficient second harmonic generation using entangled pairs~\cite{shapingsinglephoton}. Classically, the efficiency of SFG is proportional to the pump power $P_{\text{pump}}$ and the square of the crystal length $L^2.$ Commercially available nonlinear waveguides offer high normalized SFG efficiencies $\hat \eta\approx$100\%/(W$\cdot$cm$^{2})$ \cite{HCP}. Consider that the pump power is reduced so that a single photon is present per temporal mode. In this regime, the power of the input beam can be calculated as the energy of each photon divided by its coherence time, i.e. $P_{\text{pump}} = h\,\nu\,\Delta\nu/\text{tbp}$, where $\Delta\nu$ is the photon bandwidth and tbp is the time-bandwidth product. Furthermore, the bandwidth $\Delta\nu$ is limited by group velocity dispersion, and decreases linearly with the length of the crystal, i.e. $\Delta\nu= \hat \Delta\nu/L$ where $\hat \Delta\nu$ is the spectral acceptance of the crystal. The overall conversion efficiency when the full bandwidth of the crystal is used, is given by
$
\eta_{\text{SFG}}^{\text{th}} = \hat{\eta}\,\hat{\Delta\nu}\,h\nu\,L/\text{tbp}.
$
We experimentally verified that this equation holds by injecting a pair of one-photon-per-mode beams at 1557\,nm and 1563\,nm into a  2.6\,cm periodically poled lithium niobate waveguide with an acceptance bandwidth of $\hat{\Delta\nu}=300\,\text{GHz$\cdot$cm}$ and measuring the rate of 780\,nm output photons (see fig.\,\ref{fig4}).
In our experiment $\hat{\eta}=15\%/(\text{W}\cdot\text{cm}^2)$ and $\text{tbp=0.66}$, so that the expected efficiency is $\eta_{\text{SFG}}^{\text{th}} \approx 1\times 10^{-8}$. We measured an efficiency of $\eta_\text{SFG}^\text{meas}=1.2(0.2)\times 10^{-8}$. Hence, with a more appropriate commercial waveguide, 5\,cm long and $\hat{\eta}$=100\%/(W$\cdot$cm$^{2})$ \cite{HCP}, one could realistically get  $\eta_{\text{SFG}}^{\text{th}} \approx 1.5\times 10^{-7}.$ With the research device presented in Ref. \cite{Fejer} (10 cm, $\hat{\eta}=150\%/(\text{W}\cdot\text{cm}^2)$) the efficiency would increase to $\eta_{\text{SFG}}^{\text{th}} \approx 5\times 10^{-7}$. Note that the efficiency could further be improved using group velocity matching~\cite{Yu}, higher spatial confinement of the modes~\cite{Kurimura} or the use of highly nonlinear organic materials~\cite{Jaz}.\\

\paragraph{Conclusion} In conclusion, we have shown how SFG can make the entanglement resulting from entanglement swapping faithful. Despite long held preconceptions, we have demonstrated that the SFG efficiency is high enough to provide efficient, yet simpler solutions to linear-optics based protocols for the heralded production of entangled states or for the implementation of DI-QKD.\\

\paragraph{Acknowledgement} We thank M. Afzelius, J.-D. Bancal, C. Clausen, C. Osorio, H. de Riedmatten, S. Pironio, Y. Silberberg, and C. Wagenknecht for helpful discussions. We acknowledge support by the ERC-AG QORE and the Swiss NCCR {\it Quantum Photonics}.



\begin{thebibliography}{3}
\bibitem{note} For our purpose, the transmission losses and the non-unit efficiency of the single-photon detectors are not relevant.
\bibitem{footnote} This formula and the one for the fidelity are extensions of those given in Ref. \cite{Sliwa03} which take the four pair emission into account.
\end{thebibliography}
\end{document}